\documentclass[a4paper,11pt]{article}
\usepackage{pos}

\title{Worldline representation of multiloop amplitudes in quantum electrodynamics}
\author[a]{V.M. Banda Guzmán} 

\author[b]{U. M\"uller}

\author[c]{C. Nava Jacuinde}

\author*[c]{C. Schubert}

\author[c]{C.J. Servin Tomas}

\affiliation[a]{Universidad Politécnica de San Luis Potosí, Urbano Villalón 500, Colonia La Ladrillera, C.P. 78363, San Luis Potosí, San Luis Potosí, Mexico}

\affiliation[b]{Unaffiliated, Brandenburg an der Havel, Brandenburg, Germany}
\affiliation[c]{Facultad de Ciencias Físico-Matemáticas, Universidad Michoacana de San Nicolás de Hidalgo, Avenida Francisco J. Mújica, 58060 Morelia, Michoacán, Mexico} 
\emailAdd{christianschubert137@gmail.com}

\abstract{The worldline approach to quantum electrodynamics allows one to construct integral representations combining large numbers of Feynman diagrams. Here I review the state-of-the-art of a long-term effort to make this fact useful for actual multiloop calculations, such as of the scalar and spinor QED vacuum polarization functions and the electron anomalous magnetic moment. After a short historical introduction, and a discussion of the non-standard mathematical challenges involved, I focus on a recent calculation of the two-loop vacuum polarisation in scalar QED and some master integral formulas obtained in this context.}

\FullConference{Loops and Legs in Quantum Field Theory (LL2026)\\
12-17, April, 2026\\
Bayreuth, Germany\\}

\begin{document}
\maketitle

\def\nonu{\nonumber\\}
\newcommand{\intT}{\int_0^\infty \dfrac{dT}{T}\, T^{4-\frac{D}{2}}\,e^{-m^2 T}\,
\int_{(4)} 
\e^{T\,\Lambda} \;}

\newcommand{\g}[1]{G_{ #1}}
\newcommand{\dg}[1]{\dot{G}_{ #1}}
\newcommand{\ddg}[1]{\ddot{G}_{ #1}}
\newcommand{\dilog}[1]{\text{Li}_2\left(#1\right)}
\newcommand{\Log}[1]{\log\left(#1\right)}
\newcommand{\hs}{\hat{s}}
\newcommand{\hT}{\hat{t}}
\newcommand{\hu}{\hat{u}}
\newcommand{\ha}{\hat{a}}
\newcommand{\hb}{\hat{b}}
\newcommand{\hc}{\hat{c}}
\newcommand{\e}{\text{e}}
\newcommand{\Arg}{\text{Arg}}
\newcommand{\bet}[1]{\beta_{\hat{#1}}}

\newcommand{\lnbeta}[1]{\ln\left( \dfrac{\beta_{#1}-1}{\beta_{#1}+1} \right)}

\newcommand{\bifmm}[3]{\bi{ -\dfrac{\beta_{#1}}{2},-\dfrac{\beta_{#2#3}}{2} }}
\newcommand{\bifmp}[3]{\bi{ -\dfrac{\beta_{#1}}{2}, \dfrac{\beta_{#2#3}}{2} }}
\newcommand{\bifpp}[3]{\bi{ \dfrac{\beta_{#1}}{2}, \dfrac{\beta_{#2#3}}{2} }}
\newcommand{\bifpm}[3]{\bi{ \dfrac{\beta_{#1}}{2}, -\dfrac{\beta_{#2#3}}{2} }}

\def\ddel{{}^\bullet\! \Delta}
\def\deld{\Delta^{\hskip -.5mm \bullet}}
\def\dddel{{}^{\bullet \bullet} \! \Delta}
\def\ddeld{{}^{\bullet}\! \Delta^{\hskip -.5mm \bullet}}
\def\deldd{\Delta^{\hskip -.5mm \bullet \bullet}}
\def\epsk#1#2{\varepsilon_{#1}\cdot k_{#2}}
\def\epseps#1#2{\varepsilon_{#1}\cdot\varepsilon_{#2}}

\newcommand{\Ascr}{\mathscr{A}}
\newcommand{\Dscr}{\mathscr{D}}
\newcommand{\Mscr}{\mathscr{M}}
\newcommand{\Wscr}{\mathscr{W}}
\newcommand{\OWscr}{\mathscr{OW}}

\newcommand{\Acal}{\mathcal{A}}
\newcommand{\Gcal}{\mathcal{G}}
\newcommand{\Dcal}{\mathcal{D}}
\newcommand{\Mcal}{\mathcal{M}}

\newcommand{\jhat}[1]{\hspace{0.3em}\widehat{\hspace{-0.4em}#1\hspace{-0.4em}}\hspace{0.4em}}
\newcommand{\bone}{1\!\!1}

\definecolor{green}{rgb}{0.2, 0.7, 0.3}
\def\green{\color{green}}
\def\blue{\color{blue}}
\def\red{\color{red}}
\def\black{\color{black}}
%
\def\cosech{\rm cosech}
\def\sech{\rm sech}
\def\coth{\rm coth}
\def\tanh{\rm tanh}
\def\half{{1\over 2}}
\def\third{{1\over3}}
\def\fourth{{1\over4}}
\def\fifth{{1\over5}}
\def\sixth{{1\over6}}
\def\seventh{{1\over7}}
\def\eigth{{1\over8}}
\def\ninth{{1\over9}}
\def\tenth{{1\over10}}
\def\bN{\mathop{\bf N}}
\def\R{{\rm I\!R}}
\def\Eins{{\mathchoice {\rm 1\mskip-4mu l} {\rm 1\mskip-4mu l}
{\rm 1\mskip-4.5mu l} {\rm 1\mskip-5mu l}}}
\def\Z{{\mathchoice {\hbox{$\sf\textstyle Z\kern-0.4em Z$}}
{\hbox{$\sf\textstyle Z\kern-0.4em Z$}}
{\hbox{$\sf\scriptstyle Z\kern-0.3em Z$}}
{\hbox{$\sf\scriptscriptstyle Z\kern-0.2em Z$}}}}
\def\abs#1{\left| #1\right|}
\def\com#1#2{
        \left[#1, #2\right]}
\def\square{\kern1pt\vbox{\hrule height 1.2pt\hbox{\vrule width 1.2pt
   \hskip 3pt\vbox{\vskip 6pt}\hskip 3pt\vrule width 0.6pt}
   \hrule height 0.6pt}\kern1pt}
      \def\boxop{{\raise-.25ex\hbox{\square}}}
\def\contract{\makebox[1.2em][c]{
        \mbox{\rule{.6em}{.01truein}\rule{.01truein}{.6em}}}}
\def\ltap{\ \raisebox{-.4ex}{\rlap{$\sim$}} \raisebox{.4ex}{$<$}\ }
\def\gtap{\ \raisebox{-.4ex}{\rlap{$\sim$}} \raisebox{.4ex}{$>$}\ }
\def\mn{{\mu\nu}}
\def\rs{{\rho\sigma}}
\newcommand{\Det}{{\rm Det}}
\def\Tr{{\rm Tr}\,}
\def\tr{{\rm tr}\,}
\def\sumij{\sum_{i<j}}
\def\e{\,{\rm e}}
\def\non{\nonumber\\}
\def\br{{\bf r}}
\def\bp{{\bf p}}
\def\bx{{\bf x}}
\def\by{{\bf y}}
\def\brhat{{\bf \hat r}}
\def\bv{{\bf v}}
\def\ba{{\bf a}}
\def\bE{{\bf E}}
\def\bB{{\bf B}}
\def\bA{{\bf A}}
\def\pa{\partial}
\def\dA{\partial^2}
\def\ddx{{d\over dx}}
\def\ddt{{d\over dt}}
\def\der#1#2{{d #1\over d#2}}
\def\lie{\hbox{\it \$}} 
\def\partder#1#2{{\partial #1\over\partial #2}}
\def\secder#1#2#3{{\partial^2 #1\over\partial #2 \partial #3}}
\def\kinq{{1\over 4}\dot q^2}
\def\kinb{{1\over 4}\dot x^2}
%
\def\bef{\begin{frame}}
\def\ef{\end{frame}}
\def\be{\begin{equation}}
\def\ee{\end{equation}\noindent}
\def\bear{\begin{eqnarray}}
\def\ear{\end{eqnarray}\noindent}
\def\bec{\blue\begin{equation}}
\def\eec{\end{equation}\black\noindent}
\def\bearc{\blue\begin{eqnarray}}
\def\earc{\end{eqnarray}\black\noindent}
\def\benn{\begin{enumerate}}
\def\enn{\end{enumerate}}
\def\veject{\vfill\eject}
\def\ven{\vfill\eject\noindent}
%
\def\eq#1{{eq. (\ref{#1})}}
\def\eqs#1#2{{eqs. (\ref{#1}) -- (\ref{#2})}}
%
\def\totint{\int_{-\infty}^{\infty}}
\def\posint{\int_0^{\infty}}
\def\negint{\int_{-\infty}^0}
\def\pint{{\dps\int}{dp_i\over {(2\pi)}^d}}
%
\newcommand{\GeV}{\mbox{GeV}}
\def\FFdual{F\cdot\tilde F}
\def\bra#1{\langle #1 |}
\def\ket#1{| #1 \rangle}
\def\braket#1#2{\langle {#1} \mid {#2} \rangle}
\def\vev#1{\langle #1 \rangle}
\def\rightvac{\mid 0\rangle}
\def\leftvac{\langle 0\mid}
\def\ihbar{{i\over\hbar}}
\def\ge{\hbox{$\gamma_1$}}
\def\gz{\hbox{$\gamma_2$}}
\def\gd{\hbox{$\gamma_3$}}
\def\go{\hbox{$\gamma_1$}}
\def\gt{\hbox{\$\gamma_2$}}
\def\gth{\hbox{$\gamma_3$}} 
\def\gf{\hbox{$\gamma_5\;$}}
\def\slash#1{#1\!\!\!\raise.15ex\hbox {/}}
\newcommand{\slD}{\,\raise.15ex\hbox{$/$}\kern-.27em\hbox{$\!\!\!D$}}
\newcommand{\slpartial}{\raise.15ex\hbox{$/$}\kern-.57em\hbox{$\partial$}}

\newcommand{\PP}{\cal P}
\newcommand{\G}{{\cal G}}
\newcommand{\nc}{\newcommand}
\newcommand{\Fkala}{F_{\kappak_i\cdot k_j}}
\newcommand{\Fkanu}{F_{\kappa\nu}}
\newcommand{\Flaka}{F_{k_i\cdot k_j\kappa}}
\newcommand{\Flamu}{F_{k_i\cdot k_j\mu}}
\newcommand{\Fmunu}{F_{\mu\nu}}
\newcommand{\Fnumu}{F_{\nu\mu}}
\newcommand{\Fnuka}{F_{\nu\kappa}}
\newcommand{\Fmuka}{F_{\mu\kappa}}
\newcommand{\Fkalamu}{F_{\kappak_i\cdot k_j\mu}}
\newcommand{\Flamunu}{F_{k_i\cdot k_j\mu\nu}}
\newcommand{\Flanumu}{F_{k_i\cdot k_j\nu\mu}}
\newcommand{\Fkamula}{F_{\kappa\muk_i\cdot k_j}}
\newcommand{\Fkanumu}{F_{\kappa\nu\mu}}
\newcommand{\Fmulaka}{F_{\muk_i\cdot k_j\kappa}}
\newcommand{\Fmulanu}{F_{\muk_i\cdot k_j\nu}}
\newcommand{\Fmunuka}{F_{\mu\nu\kappa}}
\newcommand{\Fkalamunu}{F_{\kappak_i\cdot k_j\mu\nu}}
\newcommand{\Flakanumu}{F_{k_i\cdot k_j\kappa\nu\mu}}

\nc{\spa}[3]{\left\langle#1\,#3\right\rangle}
\nc{\spb}[3]{\left[#1\,#3\right]}
\nc{\ksl}{\not{\hbox{\kern-2.3pt $k$}}}
\nc{\hf}{\textstyle{1\over2}}
\nc{\pol}{\varepsilon}
\nc{\tq}{{\tilde q}}
\nc{\esl}{\not{\hbox{\kern-2.3pt $\pol$}}}
\newcommand{\cL}{\cal L}
\newcommand{\D}{\cal D}
\newcommand{\Dhalf}{{D\over 2}}
\def\eps{\epsilon}
\def\epshalf{{\epsilon\over 2}}
\def\lag{( -\partial^2 + V)}
\def\freeexp{{\rm e}^{-\int_0^Td\tau {1\over 4}\dot x^2}}
\def\kinb{{1\over 4}\dot x^2}
\def\kinf{{1\over 2}\psi\dot\psi}
\def\expk{{\rm exp}\biggl[\,\sum_{i<j=1}^4 G_{Bij}k_i\cdot k_j\biggr]}
\def\expp{{\rm exp}\biggl[\,\sum_{i<j=1}^4 G_{Bij}p_i\cdot p_j\biggr]}
\def\expshort{{\e}^{\half G_{Bij}k_i\cdot k_j}}
\def\expabb{{\e}^{(\cdot )}}
\def\epseps#1#2{\varepsilon_{#1}\cdot \varepsilon_{#2}}
\def\epsk#1#2{\varepsilon_{#1}\cdot k_{#2}}
\def\kk#1#2{k_{#1}\cdot k_{#2}}
\def\G#1#2{G_{B#1#2}}
\def\Gp#1#2{{\dot G_{B#1#2}}}
\def\GF#1#2{G_{F#1#2}}
\def\Dab{{(x_a-x_b)}}
\def\Dsq{{({(x_a-x_b)}^2)}}
\def\PITD{{(4\pi T)}^{-{D\over 2}}}
\def\4piTD{{(4\pi T)}^{-{D\over 2}}}
\def\4piT4{{(4\pi T)}^{-2}}
\def\TintmD{{\dps\int_{0}^{\infty}}{dT\over T}\,e^{-m^2T}
    {(4\pi T)}^{-{D\over 2}}}
\def\Tintm4{{\dps\int_{0}^{\infty}}{dT\over T}\,e^{-m^2T}
    {(4\pi T)}^{-2}}
\def\Tintm{{\dps\int_{0}^{\infty}}{dT\over T}\,e^{-m^2T}}
\def\Tint{{\dps\int_{0}^{\infty}}{dT\over T}}
\def\np{n_{+}}
\def\nm{n_{-}}
\def\Np{N_{+}}
\def\Nm{N_{-}}
\newcommand{\slG}{{{\dot G}\!\!\!\! \raise.15ex\hbox {/}}}
\newcommand{\Gd}{{\dot G}}
\newcommand{\Gund}{{\underline{\dot G}}}
\newcommand{\Gdd}{{\ddot G}}
\def\GBd12{{\dot G}_{B12}}
\def\Dx{\dps\int{\cal D}x}
\def\Dy{\dps\int{\cal D}y}
\def\Dpsi{\dps\int{\cal D}\psi}
\def\dint#1{\int\!\!\!\!\!\int\limits_{\!\!#1}}
\def\ddtau{{d\over d\tau}}
\def\ie{\hbox{$\textstyle{\int_1}$}}
\def\iz{\hbox{$\textstyle{\int_2}$}}
\def\id{\hbox{$\textstyle{\int_3}$}}
\def\ldop{\hbox{$\lbrace\mskip -4.5mu\mid$}}
\def\rdop{\hbox{$\mid\mskip -4.3mu\rbrace$}}
%
\newcommand{\1}{{\'\i}}
\newcommand{\no}{\noindent}
\def\non{\nonumber}
\def\dps{\displaystyle}
\def\sy{\scriptscriptstyle}
\def\sy{\scriptscriptstyle}

\section{QED perturbation theory in the worldline formalism}

At around the same time when Feynman developed the modern approach to perturbation theory in quantum field theory,
he also invented the ``worldline'' representation of the QED S-matrix \cite{feynman1950,feynman1951}. In diagrammatic language,
in this approach electron lines and loops are represented by relativistic particle path integrals, and interconnected by free photon
propagators in all possible ways. Although completely equivalent to the standard Feynman diagram approach, this representation
has the potential to drastically curb the usual proliferation of terms in higher orders of perturbation theory, since a priori
it does not require fixing an ordering for the photon legs along a line or loop, allowing one to combine large classes of Feynman diagrams.

\begin{figure}[htbp]
\begin{center}
 \includegraphics[width=0.35\textwidth]{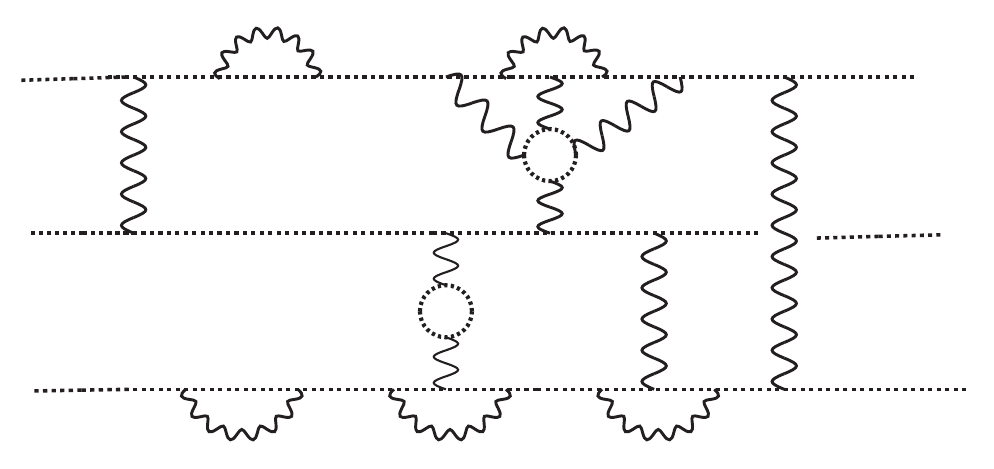}
\caption{Photon-ordered vs. photon-unordered diagrams in QED.}
\label{fig-QEDSmatrix}
\end{center}
\end{figure}

For example, the diagram shown in Fig. \ref{fig-QEDSmatrix} usually would be read as a single Feynman diagram, but in the worldline
path-integral approach naturally appears combined with several thousand others, that differ only in the ordering of some photon legs
along one of the electron lines or loops. 

The worldline representation is usually derived in terms of its natural building blocks, the one-loop
$N$-photon amplitudes and the fermion propagator dressed with $N$ photons (for details, see \cite{41}).
In scalar QED, the
former can be obtained from Feynman's worldline representation of the one-loop effective action,
\bear
\Gamma [A] &=&- {\rm \Tr}{\rm ln} \Bigl\lbrack -(\partial + ie A)^2+m^2\Bigr\rbrack  =
 \int_0^{\infty} \frac{dT}{T} \,{\rm Tr} \, {\rm exp}\Bigl\lbrack - T (  -(\partial + ie A)^2+m^2)\Bigr\rbrack  \nonumber\\
&=&
\int_0^{\infty}
\frac{dT}{T}\,
\e^{-m^2T}
\int_{x(0)=x(T)}
{\cal D}x(\tau)\,
e^{-\int_0^T d\tau \bigl(\kinb +ie\dot x\cdot A(x(\tau))\bigr)}
\, .
\label{wlpi}
\ear
Expanding the field in $N$ plane waves,
\bear
A^{\mu}(x(\tau)) = \sum_{i=1}^N \,\varepsilon_i^{\mu}\e^{ik_i\cdot x(\tau)}
\label{Apw}
\ear
and collecting all terms involving each photon polarization $\varepsilon_1,\ldots,\varepsilon_N$ linearly, 
one gets the one-loop $N$ - photon amplitudes, already summed over all permutations. 
In Feynman's original formalism, the transition to spinor QED is then made by inserting, under the path integral,
the {\it spin factor} ${\rm Spin}[x,A]$,
\bear
{\rm Spin}[x,A] = {\rm tr}_{\Gamma} {\cal P}
\exp\Biggl[{
\frac{i}{4}
e\,[\gamma^{\mu},\gamma^{\nu}]
\int_0^Td\tau F_{\mu\nu}(x(\tau))}\Biggr]
\ear
where ${\rm tr}_{\Gamma}$ denotes the Dirac trace, and $\cal P$ the path-ordering operator
(a global factor of $-\half$ needs also to be supplied to account for the change in statistics and
degrees of freedom). 
In this form it is still used today for numerical purposes, while for analytical calculations one usually
prefers Fradkin's more sophisticated representation of the spin factor in terms of an auxiliary 
Grassmann path integral \cite{fradkin66},
\bear
{\rm Spin}[x,A]  =
\int {\cal D}\psi(\tau)
\,
\exp 
\Biggl\lbrack
-\int_0^Td\tau
\Biggl(
\half \psi\cdot \dot\psi -ie \psi^{\mu}F_{\mn}\psi^{\nu}
\Biggr)
\Biggr\rbrack
\, .
\non
\ear
Here $\psi^\mu(\tau)$ is an anticommuting Lorentz vector obeying antiperiodic boundary conditions,
\bear
\psi(\tau_1)\psi(\tau_2) &=& - \psi(\tau_2)\psi(\tau_1) \, ,\quad 
\psi(T) = - \psi(0)\, .
\ear
The main point of Fradkin's approach is to replace the path-ordered exponential by an ordinary one.

\section{String-inspired treatment of the worldline path integral}

As was first  observed by Polyakov \cite{polbook}, for the plane-wave background \eqref{Apw} the worldline path integral
\eqref{wlpi} becomes gaussian, and therefore can, like the analogous Polyakov path integral of string theory,
be performed by a standard completing-the-square procedure using worldline correlators adapted to the
boundary conditions. The basic two-point correlators are 
\bear
\langle x^{\mu}(\tau_1)x^{\nu}(\tau_2) \rangle
&=&
-G(\tau_1,\tau_2)\, \delta^{\mu\nu} \, , \quad
G(\tau_1,\tau_2) = \vert \tau_1 -\tau_2\vert - \frac{1}{T} \Bigl(\tau_1 -\tau_2\Bigr)^2 \, , \non\\
\langle \psi^{\mu}(\tau_1)\psi^{\nu}(\tau_2)\rangle
&=&
\half
G_F(\tau_1,\tau_2)\, \delta^{\mu\nu} \, , \quad
G_F(\tau_1,\tau_2) = {\rm sgn}(\tau_1 - \tau_2) \, .
\nonumber\\
\label{defGGF}
\ear
This leads to  compact master formulas for the one-loop $N$-photon amplitudes
in scalar and spinor QED. For the scalar loop one finds \cite{polbook,berkosNPB,strassler1},
\begin{eqnarray}
\Gamma_{\rm scal}
[k_1,\varepsilon_1;\ldots;k_N,\varepsilon_N]
&=&
{(-ie)}^N
{\dps\int_{0}^{\infty}}{dT\over T}
{(4\pi T)}^{-{D\over 2}}
e^{-m^2T}
\prod_{i=1}^N \int_0^T 
d\tau_i
\nonumber\\
&&\hspace{-8pt}
\times
\exp\biggl\lbrace\sum_{i,j=1}^N 
\bigl\lbrack \half G_{ij} k_i\cdot k_j
+i\dot G_{ij}k_i\cdot\varepsilon_j 
+\half\ddot G_{ij}\varepsilon_i\cdot\varepsilon_j
\bigr\rbrack\biggr\rbrace
\mid_{{\rm lin}(\varepsilon_1,\ldots,\varepsilon_N)}
\nonumber\\
\label{master}
\end{eqnarray}
\no
where $\tau_i$  parametrizes the position of photon $i$ along the scalar loop, and $\mid_{{\rm lin}(\varepsilon_1,\ldots,\varepsilon_N)}$
denotes the projection on the terms linear in each photon polarization vector. Besides the worldline Green's function $G$ also its first and
second derivatives appear,
\bear
\dot G(\tau_1,\tau_2) &=& {\rm sgn}(\tau_1 - \tau_2)
- 2 {{(\tau_1 - \tau_2)}\over T}\, , \quad
\ddot G(\tau_1,\tau_2)
= 2 {\delta}(\tau_1 - \tau_2)
- {2\over T}\, .
\ear
\no
Introducing further the worldline superfield $X^{\mu} \equiv x^{\mu} + \sqrt 2\,\theta\psi^{\mu}$, 
a similar master formula can be written for the spinor loop \cite{polbook}:
\begin{eqnarray}
\Gamma_{\rm spin}
[k_1,\varepsilon_1;\ldots;k_N,\varepsilon_N]
&=&
-2
{(-ie)}^N
{\dps\int_{0}^{\infty}}{dT\over T}
{(4\pi T)}^{-{D\over 2}}e^{-m^2T}
\nonumber\\&&
\!\!\!\!\!\!\!\!\!\!\!\!\!\!\!\!\!\!\!\!
\!\!\!\!\!\!\!\!\!\!\!
\!\!\!\!\!\!\!\!\!\!\!\!\!\!\!\!\!\!\!\!
\!\!\!\!\!\!\!\!\!\!\!\!\!\!\!\!\!\!\!\!
\times
\prod_{i=1}^N \int_0^T 
d\tau_i
\int
d\theta_i
\exp\biggl\lbrace
\sum_{i,j=1}^N
\Biggl\lbrack
\half\hat G_{ij} k_i\cdot k_j
+iD_i\hat G_{ij}\varepsilon_i\cdot k_j
+\half D_iD_j\hat G_{ij}\varepsilon_i\cdot\varepsilon_j\Biggr]
\biggr\rbrace
\mid_{\varepsilon_1\ldots\varepsilon_N}
\nonumber
\label{supermaster}
\end{eqnarray}
\no
where the  $\theta_i$  are Grassmann variables and

\bear
D_i = \partder{}{\theta}_i - \theta_i \partder{}{\tau_i}, \quad
\hat G_{12} = G_{12} + \theta_1\theta_2G_{F12}\, .
\nonumber
\ear

\section{Comparison with Feynman diagrams}

The integral representations following from the scalar QED master formula \eqref{master} can,
after the restriction to any fixed ordering of the $N$ photon legs, be identified with the
ones obtained in the standard formalism using Feynman-Schwinger parameters \cite{berdun}
(the seagull vertex is hidden in the delta function contained in $\ddot G(\tau_1,\tau_2)$). 

\begin{figure}[h]
\begin{center}
\includegraphics[scale=0.45]{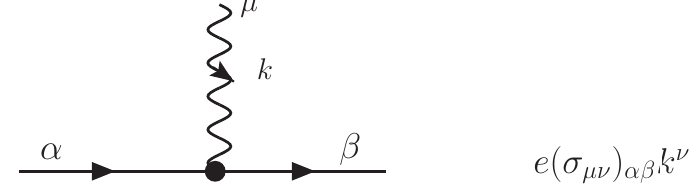}
\caption{Spin-induced vertex of the second-order formalism ($\sigma^\mn = \half [\gamma^\mu,\gamma^\nu ]$).}
\label{fig-spinvertex}
\end{center}
\end{figure}

To the contrary, the spinor QED master formula \eqref{supermaster} gives parameter integrals
corresponding not to the usual first-order Dirac formalism, but to the second-order one \cite{hostler,morgan} based on 
the identity
\bear
 ({\slash p}+e\slash A )^2 =
 -(\partial + ie A)^2 -{i\over 2}e\, \sigma^{\mn}F_{\mn} \, .
 \ear
Its Feynman rules are (up to global factors) the ones of scalar QED with a third vertex due to the spin factor, shown in Fig. \ref{fig-spinvertex}.

Despite of this principal equivalence, the worldline representation has two advantages over the Feynman diagrammatic one:

\benn

\item
 It avoids the break-up of scalar/fermion lines or loops into individual propagators.
This is important for external-field problems, where usually individual propagators have already a complicated structure.

\item

It yields the complete $ N$-photon amplitudes, with no need to sum over photon permutations.

\enn

\begin{figure}[htbp]
\begin{center}
\includegraphics[scale=0.45]{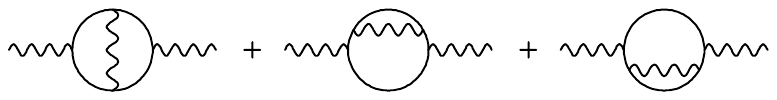}
\vspace{-240pt}
\caption{Two-loop photon propagator in spinor QED.}
\label{fig-2loopvpdiag}
\end{center}
\end{figure}

The latter fact may not seem very relevant at the one-loop level, but becomes interesting when the 
master formulas \eqref{master}, \eqref{supermaster}, which are valid off-shell, are used as building blocks
for the construction of multi-loop amplitudes. 
For example, from the four-photon amplitude we can construct the two-loop photon propagator (Fig. \ref{fig-2loopvpdiag}).
\vspace{-110pt}

\vspace{50pt}

The summation over all permutations of all photon legs in the master formula 
now has the effect of generating a sum of diagrams that involve two different 
topologies. 

\begin{figure}[htbp]
\begin{center}
\includegraphics[scale=0.35]{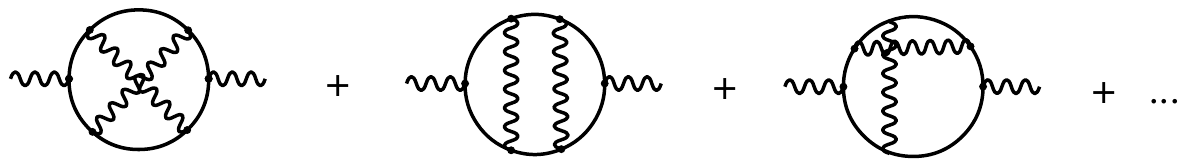}
\vspace{-130pt}
\caption{Three-loop quenched photon propagator in spinor QED.}
\label{fig-3loopbetadiag}
\end{center}
\end{figure}

Similarly, from the one-loop six-photon amplitude we can, by sewing
two pairs of photon legs, get the three-loop quenched propagator (Fig. \ref{fig-3loopbetadiag}).

\vspace{-130pt}

\noindent

Thus we get, in one single integral, Feynman diagrams of several different topologies,
and appearing with the correct multiplicities.
And, as will be familiar to many in this audience, this type of QED amplitudes
are well-known for particularly extensive cancellations between Feynman diagrams 
(see \cite{brdekr} and refs. therein). 

\section{Incorporation of background fields}

The master formulas \eqref{master}, \eqref{supermaster} can straightforwardly be generalized to the inclusion of an external constant field $F_\mn$
\cite{shaisultanov,18}. This requires

\benn

\item
Changing the worldline Green's functions $ G, G_F$ to field-dependent ones $ {\cal G}_B, {\cal G}_F$, 

\begin{eqnarray}
G(\tau_1,\tau_2) &\to &
{\cal G}_{B}(\tau_1,\tau_2) = \frac{T}{2{\cal Z}^2}
\biggl({{\cal Z}\over{{\rm sin}{\cal Z}}}
\,{\rm e}^{-i{\cal Z}\dot G_{12}}
+i{\cal Z}\dot G_{12} -1\biggr)
\, ,
\nonumber\\
G_F(\tau_1,\tau_2) &\to &
{\cal G}_{F}(\tau_1,\tau_2) =
G_{F12}
{{\rm e}^{-i{\cal Z}\dot G_{12}}\over {\rm cos}{\cal Z}}
\, ,
\nonumber\\
\end{eqnarray}
\noindent
where $ {\cal Z}_{\mu\nu} \equiv eF_{\mu\nu}T$.

\item
Adding global determinant factors

\bear
{\rm det}^{{1\over 2}}\biggl[\frac{\cal Z}{{\rm sin}{\cal Z}}\biggr] \quad { ({\rm Scalar\,QED})} \, ,\quad
{\rm det}^{{1\over 2}}\biggl[\frac{\cal Z}{{\rm tan}{\cal Z}}\biggr] \quad { ({\rm Spinor\,QED})}\, . 
\ear

\enn

Similar generalizations exist to  plane-wave backgrounds \cite{ildtor,141}
and  mixed constant-field/plane-wave backgrounds \cite{154}.

\section{Worldline integration}

Comparing the parameter integrals generated by the worldline representation 
with the ones following from the standard Feynman-Schwinger parameter approach,
it becomes immediately apparent that the ability of the worldline formalism to
combine into single integrals many Feynman diagrams of different topologies,
and often even different numbers of independent parameter integrations,
is due to the appearance of absolute value, sign and delta functions in 
$G,\dot G,\ddot G$ and $G_F$. 
However, for this property of the worldline integrals to become truly useful we have to learn how to perform them ``in one go'', that is, without splitting them into sectors with a fixed ordering of the photon legs (which would bring us back to individual Feynman diagrams). In a numerical integration this is usually not a serious complication, but for
analytical purposes one has to confront here a non-standard integration problem
for which existing mathematics seems of little help. 

Returning to the one-loop photon amplitudes, using that $G_{Fij}$'s can always be eliminated by
\bear
G_{Fij}G_{Fjk}G_{Fki} = - (\dot G_{ij} + \dot G_{jk} + \dot G_{ki})  
\ear
 the most general integral that one encounters in their calculation in the worldline formalism is of the form
\bear
\int_0^1du_1du_2 \cdots du_N \, {\rm Pol} (\dot G_{ij}) \,\e^{\sum_{i<j=1}^N \lambda_{ij}^2 G_{ij} }
\nonumber
\ear
with arbitrary $ N$ and polynomial ${\rm Pol}(\dot G_{ij})$, where (after a rescaling to the unit circle)
\bear
G_{ij} = |u_i-u_j| - (u_i-u_j)^2\, , \quad
\dot G_{ij} = {\rm sgn}(u_i-u_j) - 2(u_i-u_j)
\, .
\ear
Over the years, much effort 
\cite{41,135,156,mischa,160,victor} has gone into building tables of integrals that allow one
to do such ``circular integrals'' without decomposing the integrand into ordered sectors. 
A nice example are the ``chain integrals''
\begin{eqnarray}
\int_0^1 du_2\ldots du_n
\dot G_{12}\dot G_{23}\ldots\dot G_{n(n+1)} 
&=& -{2^n\over n!}B_n(\vert u_1-u_{n+1}\vert)
{\rm sign}^n(u_1-u_{n+1})
\, ,
\non\\
\int_0^1 du_2\ldots du_n
{G_F}_{12}{G_F}_{23}\ldots{G_F}_{n(n+1)} 
&=& {2^{n-1}\over{(n-1)!}}
E_{n-1}(\vert u_1-u_{n+1}\vert)
{\rm sign}^n(u_1-u_{n+1})
\, ,
\nonumber\\
\label{chainmasterA}
\end{eqnarray}
\no
where $B_n(x)$ and $ E_n(x)$  are the Bernoulli and Euler polynomials.
These are the only integrals needed 
for the calculation of the  low-energy limit  of the QED $ N$ - photon amplitudes
\cite{51}. 

A much more general formula was obtained in \cite{135}. For any number 
of variables $n$, and integer coefficients $k_1,\ldots, k_n$,
\bear
\int_0^1 du\,
\dot G(u,u_1)^{k_1}
\dot G(u,u_2)^{k_2}
\cdots
\dot G(u,u_n)^{k_n}
&=&
{1\over 2n}
\sum_{i=1}^n\,
\prod_{j\ne i}
\sum_{l_j=0}^{k_j}
{k_j\choose l_j}
\dot G_{ij}^{k_j-l_j}
\sum_{l_i=0}^{k_i}{k_i\choose l_i}
\non\\&&\hspace{-180pt}\times
{(-1)^{\sum_{j=1}^n l_j}
\over (1+\sum_{j=1}^n l_j)n^{\sum_{j=1}^n l_j}}
\biggr\lbrace
\Bigl( \sum_{j\ne i}\dot G_{ij} +1 \Bigr)
^{1+\sum_{j=1}^n l_j}
- (-1)^{k_i-l_i}
\Bigl(
\sum_{j\ne i}\dot G_{ij} -1
\Bigr)^{1+\sum_{j=1}^n l_j }
\biggr\rbrace
\, .
\nonumber\\
\ear
By recursion, this formula allows one to calculate any integral polynomial in 
$\dot G$ and $G_F$. This suffices already, for example,  
for the computation of the full heat-kernel expansion  in scalar and spinor QED.

\section{Multi-loop worldline Green's functions}

\begin{figure}[h]
\begin{center}
\includegraphics[width=0.65\textwidth]{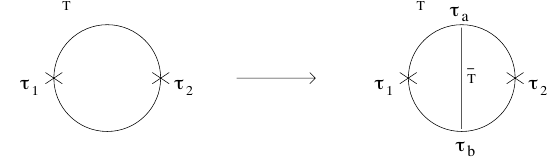}
\caption{Worldline Green's function modified by a propagator insertion.}
\label{fig-2loopG}
\end{center}
\end{figure}

The construction of higher-loop integrals by sewing of pairs of external legs
is done most directly at the parameter integral level (see, e.g., \cite{100, 137}). 
A more elegant alternative is to performing the sewing already at the path-integral level,
and to absorb the resulting internal popagators into the worldline Green's functions
\cite{8}. For a single insertion, this produces an effective two-loop Green's function
%

%
\be
G_B^{(1)}(\tau_1,\tau_2)=
G(\tau_1,\tau_2) + \half
{{[G(\tau_1,\tau_a)-G(\tau_1,\tau_b)]
[G(\tau_a,\tau_2)-G(\tau_b,\tau_2)]}
\over
{{\bar T} + G(\tau_a,\tau_b)}}
\label{defG1}
\ee
\no
where $ \bar T$ is the proper-time length of the inserted propagator, and
$ \tau_a,\tau_b$ the points on the loop between which the propagator is inserted,
see Fig. \ref{fig-2loopG}.

\vspace{2pt}

Independently of how the sewing is done, for the quenched $ l$-loop photon propagator it produces parameter integrals naturally written in the variables 

$$ G_{a_1b_1}, G_{a_2b_2},\ldots , G_{a_lb_l},C_{a_1b_1a_2b_2},\ldots , C_{a_{l-1}b_{l-1}a_lb_l}$$

\noindent
where the $ G_{a_ib_i}$ depend only on a single propagator, and the $ C_{a_ib_i a_j b_j}$ on all possible pairs of propagators. 

All this generalizes to the inclusion of constant external fields, and was used for
the calculation of two-loop Euler-Heisenberg Lagrangians scalar as well as in spinor QED, in various types of fields \cite{18,51}.

\vspace{-10pt}
\begin{figure}[h]
\begin{center}
\includegraphics[width=0.15\textwidth]{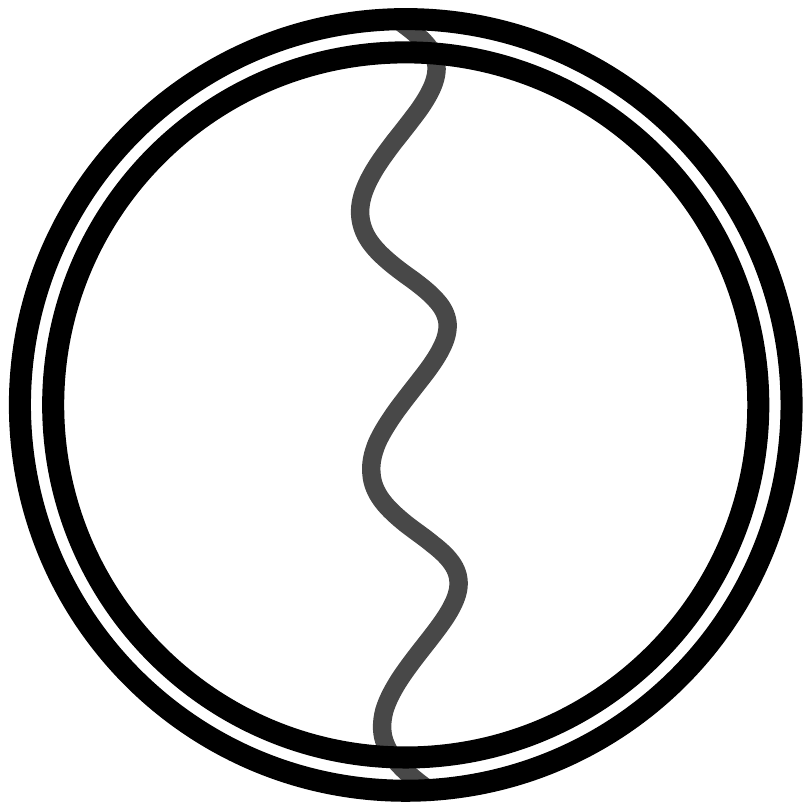}
\caption{Irreducible contribution to the 2-loop Euler-Heisenberg Lagrangian in
scalar or spinor QED.}
\label{fig-EHL1PI}
\end{center}
\end{figure}

 In particular, in \cite{51} simple explicit formulas
were found for the irreducible contributions to these Lagrangians, depicted in
Fig. \ref{fig-EHL1PI}, for the case of a (euclidean) self-dual field:
\bear
{\cal L}_{\rm scal}^{(2)}(\kappa)
&=&
\alpha \,{m^4\over (4\pi)^3}\frac{1}{\kappa^2}\left[
{3\over 2}\xi^2 (\kappa)
-\xi'(\kappa)\right]
\, ,
\\
{\cal L}_{\rm spin}^{(2)}(\kappa)
&=&
-2\alpha \,{m^4\over (4\pi)^3}\frac{1}{\kappa^2}\left[
3\xi^2 (\kappa)
-\xi'(\kappa)\right]
\, .
\ear\no
Here $ \kappa\equiv \frac{m^2}{2ef}$, where $f$ is defined by 
$F_{\mu\nu}F^{\nu\lambda} =  -f^2 \delta^{\lambda}_{\mu}$. 
$\xi(x)\equiv -x\Bigl(
\Gamma'(x)/\Gamma(x)
-\ln(x)+{1\over 2x}\Bigr)$.
Combined with the standard spinor-helicity technique, this
allowed \cite{51} to obtain explicit formulas also for the corresponding
two-loop all + helicity $N$ - photon amplitudes in the low-energy limit.

\section{Three worldline master integrals involving two propagators}

In a systematic effort to extend worldline integration to the level
of full two-loop two-point functions, we (VMB, UM and CS) 
have recently analyzed the following three families of integrals:

\benn

\item
Integrating out two internal propagators (simple):
\bear
I^m_{kl}  \equiv \int_0^1 du_1 du_2du_3du_4 \frac{C^{2m}}{G_{12}^l G_{34}^k}
\quad (k,l\leq m)\, ,
\ear

\bear
I^m_{kl} &=&
4^{m+1} 
\frac
{(\Gamma(2m-k-l+3))^2}
{\Gamma(4m-2k-2l+5)}
\frac{1}
{(2m-l+1)(2m-l+2)}
\nonumber\\ && \hspace{-20pt} \times
\biggl\lbrace
\frac{2m-2l+3}{2m-k+1}
\,{}_3F_2(1,k,l-1;2m-k+2,2m-l+3;1) \nonumber\\
&& - \frac{2m(2m-2l+1)}{(2m+1)(2m-k+2)}\,
\,{}_3F_2(1,k,l;2m-k+3,2m-l+3;1)
\biggr\rbrack
\, .
\label{Ialt}
\ear
The $ {}_3F_2$ function values are of a type that can be expressed as a
 finite sum of
values of the digamma function $\psi$.
%

\item
Integrating out the two external legs, and writing the result in terms of the
Green's function between the internal variables (more difficult):
\bear
I^{km} \equiv 
\int_{12} G_{12}^ k C^{2m} 
\quad 
(m,k \in \mathbb{N},m\geq 1)
\, ,
\ear

\bear
I^{km}
=
m(2 m - 1) \frac{\Gamma(2 m + 2 k - 1)}{\Gamma(2 m + 2 k + 3)}
\sum_{n=0}^k
(n+1)
\, 4^{m + 1-n}
 \frac{\binom{k+2m}{n} \binom{k}{n}}
{\binom{k+m-3/2}{n}\binom{k+m-1}{n}}
G_{34}^{k+2m-n}
 \, .
\nonumber\\
\label{uwe}
\ear

\item
Integrating out an internal propagator and writing the result in terms of the
Green's function between the external variables (difficult):

\bear
I^{m}_{k} \equiv \int_0^1du_a\int_0^1du_b \frac{C^{2m}}{G_{ab}^k}
\qquad (k \leq m) \, ,
\ear

\enn

\bear
I_k^m &=& 
\gamma_k^m 
\biggl\lbrace
 \ln (G_{12}) \frac{(1-\dot G_{12})^{2m}}{4^m}\Bigl\lbrack  b_+(k,m)(1-\dot G_{12})
+  b_-(k,m) (1+\dot G_{12})
\Bigr]\nonumber\\ && \!\!\!\!\!\!\!
+ 2 \sum_{j=0}^m \Bigl(b_+(k,m)d_j^m + b_-(k,m)d_{j-1}^{m-1}\Bigr)
\Bigl\lbrack
\dot G_{12} \ln \Bigl(\frac{1+\dot G_{12}}{2}\Bigr)
- 
 \sum_{n=1}^{2m-k-j}\alpha_nG_{12}^n\Bigr\rbrack 
 G_{12}^j 
\biggr\rbrace
\, ,
\label{taylorIkm}
\ear

%
\bear
\gamma_k^m &=& 
\frac{ (-1)^k 4^m}{(k-1)!(m+\half)}
\frac{(2m-k)!}{(2m-2k+2)!} 
\, ,
\\
\alpha_n&=& C_{n-1} \bigl(\psi(n-1/2) - \psi(n+1) + \ln 4\bigr) \, ,
\ear
\vspace{-20pt}
\bear
b_+(k,m) &=& - (2m+1-k)(m-1/2) \, ,\\ 
b_-(k,m) &=& -(k-1)(m+1/2) \, ,
\ear

\vspace{-20pt}

\bear
c_j^m &=&{ \binom{2m-j+1}{j} } \, ,\quad
%
%
%
d_j^m = (-1)^j (c_j^m -c_{j-1}^{m-1}) 
\ear
($C_n = \frac{1}{n+1} \binom{2n}{n}$ is the $n$th Catalan number).

\section{Two-loop vacuum polarisation in scalar QED}

\begin{figure}[h]
\begin{center}
\includegraphics[width=0.35\textwidth]{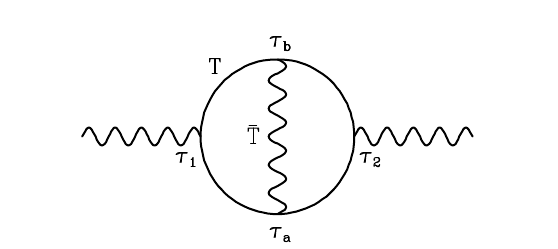}
\raisebox{20pt}{+ $\cdots$ }
\caption{Worldline parametrization of the two-loop vacuum polarization function.}
\label{fig-2loopvpint}
\end{center}

\end{figure}

Two of the authors (VMB and CS) have used the two-loop Green's function 
\eqref{defG1} to obtain the following compact integral representation for the  two-loop photon polarization function in scalar QED, parametrized according to 
Fig. \ref{fig-2loopvpint}:
\bear
\Pi^{(2)}_{\text{scal}} (k^2) &=&
\frac{e^4}{2(4\pi)^D}
 \int_0^\infty \dfrac{dT}{T^{1+\frac{D}{2}}} 
e^{-m^2T} 
\int_0^\infty d\bar T
\int_0^Td\tau_a \int_0^Td\tau_b 
 \gamma_{ab}^{D/2}
 \nonumber \\
 && \times
 \int_0^Td\tau_1 \int_0^Td\tau_2
 \, 
\e^{- \Delta k^2}
I
\, ,
\ear

\bear
I &=&
D
\Bigl(-\ddot G_{ab} + \frac{\gamma_{ab}}{2}\dot G_{ab}^2\Bigr)
\partial_1\Delta \partial_2\Delta
-
\frac{4}{\gamma_{ab}^2 C^2} \partial_a\Delta  \partial_b\Delta
\partial_1\partial_2\Delta
+
\partial_2 \Bigl\lbrack
\frac{4}{\gamma_{ab}^2 C^2} \partial_a\Delta  \partial_b\Delta
\Bigr\rbrack
\partial_1\Delta
\label{defIprimecomp}
\nonumber\\
\ear
($C=G_{1a}-G_{1b}-G_{2a}+G_{2b}, \quad
 \gamma_{ab} = (\bar T + G_{ab})^{-1},
\quad \Delta=G_{12}-\frac{\gamma_{ab}}{4}C^2)$.

Once more the challenge is to calculate this six-parameter integral without
splitting it into sectors with a fixed photon ordering. 
Expanding in $ s= - \frac{k^2}{m^2}$ and eliminating the then trivial $ T$ - integral,
\bear
\Pi^{(2)}_{\text{scal}} (k^2) &=&
\frac{e^4}{2(4\pi)^D(m^2)^{4-D}}
\sum_{n=0}^\infty \frac{\Gamma(n+4-D)}{n!} 
c_n(D)
s^n
\ear
with coefficients
\vspace{-10pt}
\bear
c_n(D) &=& \int_0^\infty d\hat T
\int_0^1du_a du_b  du_1 du_2
\, 
 \gamma_{ab}^{D/2}I\Delta^n
 \, .
 \label{In}
 \ear
 Using the binomial formula for $\Delta^n$, we can write this as

\bear
c_n(D) &=& \sum_{k=0}^n \binom{n}{k} \Bigl(-\frac{1}{4}\Bigr)^{n-k}
\int_{ab}
\int_0^\infty d\hat T
 \gamma_{ab}^{D/2+n-k}
\int_{12}
\, 
 G_{12}^k
 C^{2n-2k}
 I
 \label{In}
 \ear
 after which \eqref{uwe} can be used. Explicit calculation of the first few coefficients gives
\bear
c_0 &=& \frac{4}{3} - \frac{1}{\epsilon} \nonumber\\
c_1 &=& - \frac{41}{162} + \frac{1}{10\epsilon} \nonumber\\
c_2 &=& \frac{41}{18900} + \frac{1}{70\epsilon} \nonumber\\
c_3 &=& \frac{24287}{7938000} +\frac{1}{420 \epsilon} \nonumber\\
c_4 &=& \frac{1341383}{1571724000} + \frac{1}{2310 \epsilon} \nonumber\\
\ear
(the $1/\epsilon$ poles are still to be removed by mass renormalization). 
All these coefficients have been
confirmed by an independent calculation by  L. Tancredi and
 F. Forner 
(curiously, we could not find a result in the literature for the full two-loop vacuum polarisation function in scalar QED).

\section{Three-loop $\beta$-function coefficients}

\begin{figure}[htbp]
\begin{center}
 \includegraphics[width=0.35\textwidth]{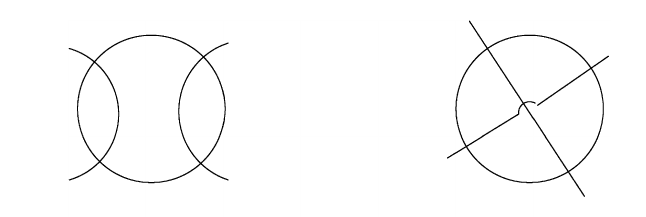}
\caption{Diagrams contributing to the $\beta$-function in $\phi^4$-theory.}
\label{fig-beta}
\end{center}
\end{figure}

The above integral formulas are also useful for the calculation of three-loop
vacuum integrals. Since those are of little physical interest in QED, let us
now switch to $\phi^4$-theory, where the diagrams shown in Fig. \ref{fig-beta},
taken at zero momentum for all the external legs, contribute to the $\beta$-function.

In \cite{41} it had already been shown that,
in the worldline formalism, the calculation of the sum of the $ 1/\epsilon$ poles
of these diagrams can be reduced to the calculation of the single integral

\begin{eqnarray}
I_{\rm reg}&=& 
\int\limits_{\hspace{4mm}1234}
\biggl[-{4\over C^2}{\rm ln}
\Bigl(1-{C^2\over 4G_{12}G_{34}}\Bigr)
-{1\over G_{12}G_{34}}\biggr]
\, .
\label{Ireg}
\end{eqnarray}

However, the actual calculation of this integral was done by decomposing it into its planar (non-crossing propagators) and
non-planar (crossing propagators) sectors $ I_{\rm reg}^P$ and $ I_{\rm reg}^{NP}$ , resulting in

\bear
I_{\rm reg}^P = 4\zeta(2)-4 \, ,\quad  I_{\rm reg}^{NP} = 12\zeta(3)  -8\zeta (2) \, .
\ear
Using the above formulas, this can now be avoided. Expanding the logarithm in \eqref{Ireg} we get
\bear
I_{\rm reg}&=& 
\sum_{n=2}^{\infty} \frac{I^{n-1}_{nn}}{n\, 4^{n-1}}
\, .
\label{Iregsum}
\end{eqnarray}
From the formula \eqref{Ialt} for $ I^m_{kl}$ one finds the special case
\bear
I^{n-1}_{nn} &=& 4^n
\Bigl\lbrack
\frac{1}{(n-1)^2} 
+\frac{ \psi'(n)}{n-\half}
\Bigr\rbrack
\ear
which reduces this more ``principled'' recalculation of $ I_{\rm reg}$ to two easy summation problems, 

\bear
\sum_{n=1}^\infty \frac{1}{n^2 (n+1)} = \zeta(2)-1 \, , \quad 
\sum_{n=1}^\infty \frac{\psi'(n+1)}{(n+\half) (n+1)} = 3\zeta(3) -2\zeta(2) \, .
\ear

\section{Worldline representation of dressed electron propagator}

While the worldline formalism has been applied to amplitudes and effective actions
involving closed fermion loops for several decades, only during the last few years
a suitable generalization was developed for problems involving open
fermion lines \cite{130,131}. 

Here the starting point is the second-order representation of the $x$-space Dirac propagator $S^{xx'}[A]$ in a Maxwell background:
 
\bear
S^{xx'}[A] &=&
\bigl[m + i\slash{D}'\bigr]
K^{xx'}[A]\,  \nonumber,\\
K^{xx'}[A] &=&
\Big\langle x'\Big|\Bigl[m^2- D_{\mu}D^{\mu} +{i\over 2}\, e \gamma^{\mu}\gamma^{\nu} F_{\mu\nu}\Bigr]^{-1} 
\Big| x \Big \rangle \nonumber\\
&=&
\int_0^{\infty}
dT\,
\e^{-m^2T}
\e^{-\fourth \frac{(x-x')^2}{T}}
\int_{q(0)=0}^{q(T)=0}
Dq\,
\e^{-\int_0^T d\tau\bigl(
\kinq
+ie\,\dot q\cdot A 
+ie \frac{x'-x}{T}\cdot A
\bigr)}
\nonumber\\
&& \times\,  2^{-\frac{D}{2}}
{\rm symb}^{-1}
\int_{\psi(0)+\psi(T)=0
\hspace{-30pt}}D\psi
\, \e^
{-\int_0^Td\tau\,
\bigl[\half\psi_{\mu}\dot\psi^{\mu}-ieF_{\mu\nu}(\psi+\eta)^{\mu}(\psi+\eta)^{\nu}\bigl]
}
\, .
\ear\no

Here $\eta^{\mu}$ is an external Grassmann Lorentz vector, and the ``symbol map''  {\it symb}  converts products of $\eta$'s 
into fully antisymmetrised products of Dirac matrices \cite{fradkin66,fragit}
\bear
{\rm symb} 
\bigl(\gamma^{\alpha_1\alpha_2\cdots\alpha_n}\bigr) \equiv 
(-i\sqrt{2})^n
\eta^{\alpha_1}\eta^{\alpha_2}\ldots\eta^{\alpha_n}
\ear
%
where $\gamma^{\alpha\beta\cdots\rho}$ denotes the totally antisymmetrised product of gamma matrices.
%
%
The dressed propagator in momentum space $K_N^{pp'} $ can be written as
\bear
	K_N^{pp'} &=&  (-ie)^N \frac{\mathfrak{K}_N^{pp'}}{\left(p'^{2} + m^{2}\right)\left(p^2 + m^{2}\right)}
\, , \quad
\mathfrak{K}_N^{pp'} =
		 A_N \Eins + B_{N\alpha\beta} \sigma^{\alpha\beta}  -i C_N\gamma_{5} \, ,
\label{defABC}
\ear\no
that is, directly in terms of the Clifford basis with only three coefficient functions
$ A_N, B_{N\alpha\beta},C_N$.
Moreover, for on-shell fermions those become algebraically dependent, 
and only $B_{N\alpha\beta}$ needs to be really calculated. 
From \cite{130,131} two master formulas can be derived for this quantity,
one using worldline supersymmetry as in \eqref{supermaster}, and
a very different one based on a spin-orbit decomposition.
We (CNJ,CS and CJST) are presently programming both master formulas
with a view on applications to nonlinear Compton scattering and
multiloop g-2. 

\section{Outlook}

We have summarized recent progress in a long-term effort to develop methods in the
worldline formalism that effectively allow one to integrate, in one go, whole classes of Feynman diagrams differing only in the ordering of photon legs. 
The  master integration formulas presented here solve this problem for the two-loop QED vacuum polarization, 
and can also be used for certain three-loop calculations involving
external legs though only at zero-momentum. We also expect them to become useful in a future calculation of the much-needed
weak-field expansion coefficients of the three-loop Euler-Heisenberg Lagrangians in scalar and spinor QED.

\end{document}